\documentclass[conference]{IEEEtran}
\IEEEoverridecommandlockouts
\usepackage{cite}
\usepackage{amsmath,amssymb,amsfonts}
\usepackage{algorithmic}
\usepackage{graphicx}
\usepackage{textcomp}
\usepackage{xcolor}

\makeatletter
\providecommand{\linebreakand}{%
  \end{@IEEEauthorhalign}%
  \hfill\mbox{}\par
  \mbox{}\hfill\begin{@IEEEauthorhalign}%
}
\makeatother

\def\BibTeX{{\rm B\kern-.05em{\sc i\kern-.025em b}\kern-.08em
    T\kern-.1667em\lower.7ex\hbox{E}\kern-.125emX}}
\begin{document}

\title{Democratizing Federated Learning with Blockchain and Multi-Task Peer Prediction}

\author{\IEEEauthorblockN{Leon Witt}
\IEEEauthorblockA{\textit{Dept. of Comp. Sci. \& Tech.} \\
\textit{Tsinghua University \& Fraunhofer HHI}\\
leonmaximilianwitt@gmail.com}
\and
\IEEEauthorblockN{Wojciech Samek}
\IEEEauthorblockA{\textit{Dept. of Elec. Eng. \& Comp. Sci.} \\
\textit{TU Berlin \& Fraunhofer HHI}\\
wojciech.samek@hhi.fraunhofer.de}
\linebreakand
\IEEEauthorblockN{Kentaroh Toyoda}
\IEEEauthorblockA{\textit{Inst. of High Performance Computing} \\
\textit{A*STAR} \\
kentaroh.toyoda@ieee.org}
\and
\IEEEauthorblockN{Dan Li}
\IEEEauthorblockA{\textit{Dept. of Comp. Sci. \& Tech.} \\
\textit{Tsinghua University}\\
tolidan@tsinghua.edu.cn}
}

\maketitle

\begin{abstract}
The synergy between Federated Learning and blockchain has been considered promising; however, the computationally intensive nature of contribution measurement conflicts with the strict computation and storage limits of blockchain systems. We propose a novel concept to decentralize the AI training process using blockchain technology and Multi-task Peer Prediction. By leveraging smart contracts and cryptocurrencies to incentivize contributions to the training process, we aim to harness the mutual benefits of AI and blockchain. We discuss the advantages and limitations of our design.
\end{abstract}


\section{Introduction}

% \todo{Find appropriate place: A promising use-case of Blockchain in the domain of AI is Federated Learning, a technique to train AI models in a decentralized fashion \cite{SP_Blockchain_IEEE, SP_MD_IEEE, SP_MD_ARXIV, SP_FL_TECHRXIV, SP_FL_MD, FLBlockchainOpportunitiesChallanges, BlockchainFLSurvey, LeonSP}. }

\begin{figure*}[t]
    \centering
    \includegraphics[width=\textwidth]{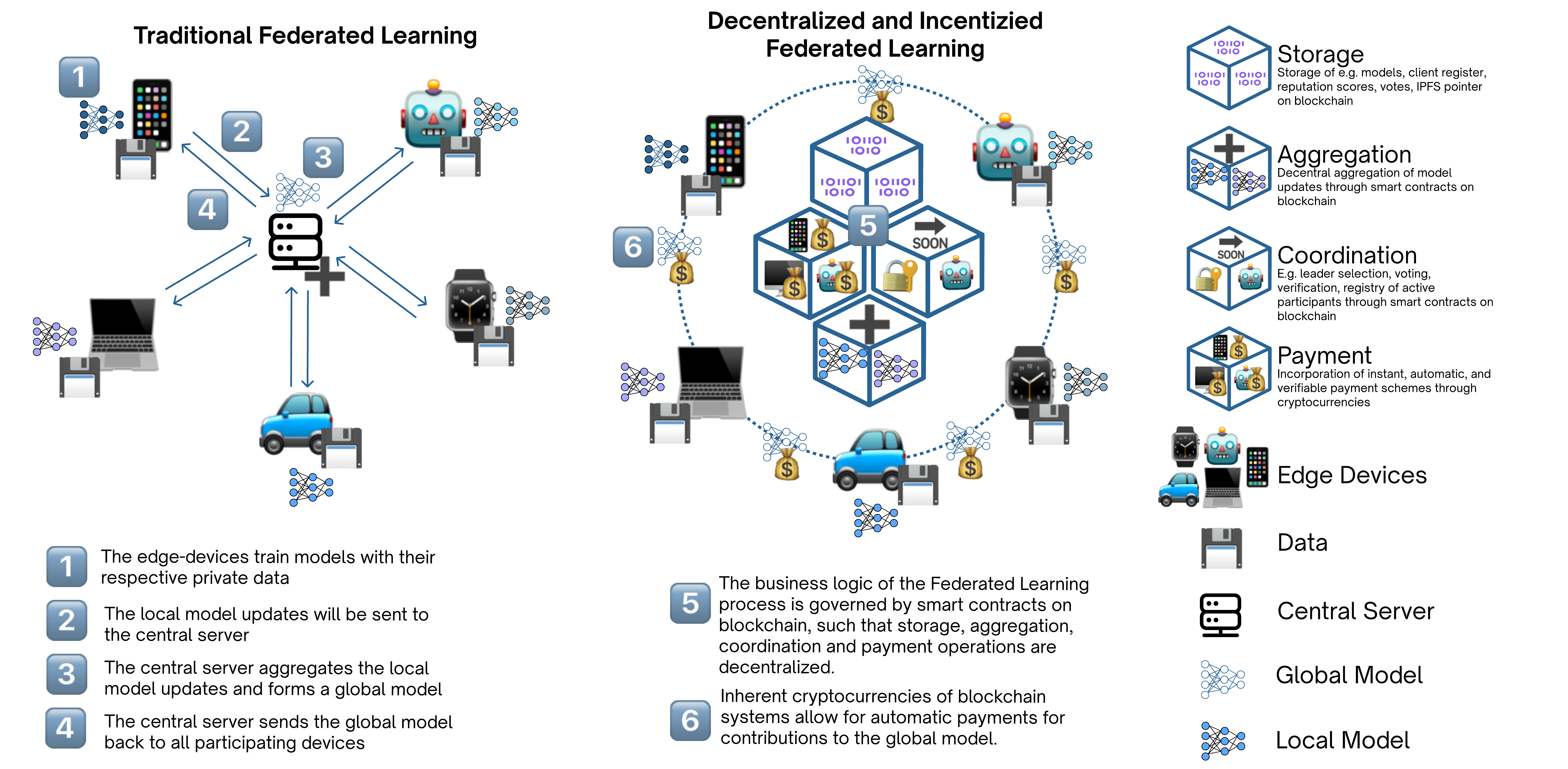}
    \caption{FL vs. Decentralized and Incentivized FLF from \cite{LeonSP}}
    \label{fig:FLvsFLFs}
\end{figure*}

Federated Learning (FL), a technique where multiple clients train an AI model locally and in parallel without data leaving the device, was pioneered by Google in 2016~\cite{BrendanMcMahan2017}. This approach (i) decentralizes the computational load across participants and  (ii) allows for privacy-preserving AI training as the training data remains on the edge devices. In this context, Federated Averaging\footnote{Other optimization algorithms for FL exist and are ongoing research~\cite{AdvancesAndOpenProblemsInFederatedLearning}, with variations like \texttt{FedBoost}\cite{Fedboost}, \texttt{FedProx}\cite{FedProx}, \texttt{FedNova}\cite{FedNova}, \texttt{FedSTC}\cite{SatTNNLS20}, and \texttt{FetchSGD}~\cite{fetchSGD} exploring improvements to the \texttt{FedAvg} algorithm.} (\texttt{FedAvg}) is an algorithm~\cite{BrendanMcMahan2017} widely adopted in FL, aiming to minimize the global model's empirical risk by aggregating local updates, represented as

\begin{equation}
\arg\min_{\boldsymbol{\theta}} \sum_{i} \frac{|S_i|}{|S|} f_{i}(\boldsymbol{\theta})
\end{equation}
where for each agent $i$, $f_{i}$ represents the loss function, $S_i$ is the set of indexes of data points on each client, and $S:=\bigcup_{i} S_i$ is the combined set of indexes of data points of all participants. The traditional Federated Learning process contains four steps and is depicted on the left of Fig.~\ref{fig:FLvsFLFs}. To scale FL Frameworks beyond entrusted entities towards mass adoption, where participants are treated equally and fairly towards genuine democratic AI training, two major challenges have to be overcome:

\begin{enumerate}
    \item \textbf{Incentivization:} Use cases where clients have valuable training data yet no interest in the AI model require incentive mechanisms through compensation. The privacy-by-design nature of FL makes it challenging to measure contributions for a fair reward distribution.
    
    \item \textbf{Decentralization:} FL, although intended to be decentralized and private, relies on a central server for aggregating gradients (FedAvg). This central server can (i) exclude clients, (ii) introduce a single point of failure, and (iii) withhold payment in environments where clients are compensated for their contributions (incentivized FL). These issues can be mitigated by replacing the central authority with blockchain technology. 
\end{enumerate}

Although the synergy between FL and blockchain has been considered to be promising \cite{witt2024blockchain}, computationally intensive processes on the blockchain, such as contribution measurement, hinders its adoption in the real world. 

In this paper, we first propose Multi-Task Peer Prediction (MTPP), a lightweight incentivization method from the domain of crowdsourcing, based on the correlation of clients’ outputs, to incentivize honest model training from clients in Federated Learning (FL). We then propose a novel framework to decentralize the AI training process using blockchain technology with MTPP. This framework leverages smart contracts and cryptocurrencies to incentivize contributions to the training process, offering a practical solution to harness the mutual benefits of AI and blockchain.

This work is structured as follows: Section~\ref{sec:ContributionMeasurement} explores contemporary methods of measuring contributions in FL. Section~\ref{sec:BlockchainInFL} examines the integration of blockchain technology within FL. Finally, Section~\ref{sec:FLF} describes a conceptual framework that combines MTPP with existing General-Purpose Blockchain Systems (GPBS) to both incentivize and decentralize FL.

\section{Contribution Measurement in FL}
\label{sec:ContributionMeasurement}

In FL, measuring individual contributions is challenging due to the privacy-preserving nature of the system, where only model updates are shared without revealing underlying data quality or distribution \cite{LeonSP, AdvancesAndOpenProblemsFL, LW1_witt2021rewardbased}. Contribution measurement methods include \textit{honest reporting} based on data volume, local accuracy, or cost, though these may encourage dishonesty due to lack of verification \cite{LW6_data1, LW11_RewardResponseGame, GTG_28, GTG_31, GTG_5, LW4_FL_IIoT, GTG_4, MH1_Chai2021, 10_GTG, EntropyLossonFullInfo, CrowdsourcingFrameworkFL, GTG_22, TXX}. \textit{Reputation systems} and majority voting offer alternative metrics, reducing the need for stringent controls, but accurately quantifying reputation remains a challenge \cite{LW5_FL_HomeAppliances_IoT, LW7_Zhang_Reputation, crowdsensing_meets_FL, CrowdsourcingFrameworkFL, KT02_Toyoda2020Access, KT01_Toyoda2019BigData}.

Direct performance comparison methods include the \textit{Leave-one-out strategy}, where the impact of removing one client's data is analyzed, although it overlooks complex data interactions \cite{leaveoneout}. The \textit{Shapley value} method accounts for all possible contributions through cooperative game theory, ensuring fairness and promoting efficiency, symmetry, and proportional rewards, though it is computationally demanding \cite{shapley1953value, shapley2, shapleyMLicml, ShapleyFLSpringer, AISTATSshapleyML, gtgShapley, ProfitAllocationFL, wang2020principled, wei2020efficient, LW9_FedCoin, MH4_Ma2021}. These methods however require a central authority, a test set to perform the measurement against and cause computational overhead \cite{GTG_AAAI}. 

\subsection{Peer Prediction}
\label{MTPP}
MTPP is a part of crowdsourcing, where a scoring rule is defined such that the optimal strategy for each client is to behave in a way the mechanism designer intends. Specifically, clients receive the most rewards for (i) exerting effort into solving the \textit{tasks} and then (ii) reporting the results of the task truthfully. The beauty of peer prediction is that it does not require knowing the ground truth to elicit said properties, as the scoring function is based on the correlations of what clients report. To understand how MTPP can be applied to FL, we need to define what constitutes a \textit{task}, \textit{signal}, and \textit{report} in the context of FL.

Examples of \textit{tasks} in FL include the classification of images \cite{LIU_CA} or finding the right parameterization of a neural network \cite{CA_HONGTAO}. For image classification, retrieving a signal means applying inference on a trained model to predict a label, where the label would be the signal. In contrast, when the task is correct parameterization, training the model to retrieve the correct parameterization would constitute retrieving a signal. Whether the client then reports the retrieved signal truthfully depends on the incentivization. An \textit{informed strategy} involves varying reports based on the signal, whereas an \textit{uninformed strategy} maintains random reports independent of the true signal. A mechanism is informed-truthful if truthful reporting ($\mathbb{I}$) always results in at least as high expected payment as any other strategy, achieving equality only when all strategies are fully informed \cite{OriginalCA}. In the context of FL, the optimal client behavior (both genuine AI training and honest reporting) should achieve a Bayesian Nash equilibrium, aligning individual incentives with the overall objectives of the FL system.

\textbf{Multi-Task Peer Prediction:} The MTPP algorithm works as follows for two clients as part of a broader group:
\begin{enumerate}
\item Assign shared tasks to ensure diverse engagement.
\item Define subsets $M_b$ as “bonus tasks” and $M_1$, $M_2$ (non-overlapping) as “penalty tasks.”
\item Payment for a bonus task is calculated by the difference in scores for bonus versus penalty tasks, with the total payment being the aggregate of all bonus tasks.
\end{enumerate}

The objective of MTPP algorithms is to define a scoring function $\mathcal{S}$, which determines how rewards should be paid based on the clients’ reports, ensuring informed-truthful behavior. Correlated Agreement \cite{OriginalCA} defines such a scoring function, based on the correlation matrix over all signals.\footnote{Please refer to \cite{OriginalCA} for further information and \cite{CA_HONGTAO,LIU_CA} for the application of CA in FL.}

% \textbf{Preliminary Definitions}: Consider two random variables $Z_1$ and $Z_2$ representing signals received by clients 1 and 2, by working on a 'task', who report $R_1$ and $R_2$, respectively. 

% methods are effective for incentivizing FL where direct measurement of contributions is challenging. MTPP methods, such as Peer Truth Serum \cite{LW1_witt2021rewardbased} and Correlated Agreement (CA) \cite{LIU_CA, CA_HONGTAO}, motivate clients to train models (informed behavior) and report truthfully in the absence of direct observation by leveraging correlations in their reports on shared tasks.

\section{Application of Blockchain in Federated Learning}
\label{sec:BlockchainInFL}

Blockchain technology appears promising in addressing several pressing issues within the context of FL:

\begin{itemize}
\item \textbf{Decentralization}: Traditional server-worker topologies in FL are susceptible to power imbalances and single points of failure. A server in such setups could potentially withhold payments or arbitrarily exclude participants. Moreover, this model does not suit scenarios where multiple stakeholders have an equal interest in jointly developing models. Blockchain technology supports a decentralized framework, eliminating the need for a central server and allowing multiple entities to cooperate as peers with equal authority.
\item \textbf{Transparency and Immutability}: Blockchain ensures that data can be added but not removed, with each transaction recorded permanently. This feature is crucial in FL, where a transparent and immutable record of rewards fosters trust among participants. Additionally, the system's auditability ensures accountability, deterring malicious activities.
\item \textbf{Cryptocurrency}: Many blockchain platforms incorporate cryptocurrency functionalities, enabling the integration of payment systems within the smart contracts. This allows for immediate, automatic, and transparent compensation of workers based on pre-defined rules within the FL reward mechanisms.
\end{itemize}

Consequently, significant research efforts have been dedicated to developing frameworks that utilize blockchain to decentralize FL \cite{SP_Blockchain_IEEE, SP_MD_IEEE, SP_MD_ARXIV, SP_FL_TECHRXIV, SP_FL_MD, FLBlockchainOpportunitiesChallanges, LeonSP}. Nevertheless, the high computational demands of model training and substantial storage requirements present considerable challenges. As of now, none of the blockchain-integrated FL frameworks are ready for production deployment \cite{LeonSP}. Blockchain's capabilities potentially extend to several complementary aspects of FL, including aggregation, payment, coordination, and storage:

\begin{enumerate}
\item \textbf{Aggregation}: In this approach, each client $i$ transmits their model parameters $\theta_i$ directly to the blockchain rather than to a central server. The aggregation of model parameters is then conducted on the blockchain via smart contracts. This method enhances robustness by eliminating single-point failures and ensures transparency and auditability in contributions\cite{LW11_RewardResponseGame, LW2_Weng2021, MH5_Lei2021, MH9_He2021, KT01_Toyoda2019BigData, KT02_Toyoda2020Access}. Although this approach results in considerable computational and storage overhead that scales with $\mathcal{O}(tnm)$, techniques such as Federated Distillation \cite{LW1_witt2021rewardbased} and two-layer BC frameworks \cite{MH5_Lei2021}, where computation layer is separated from the consensus layer, can mitigate these costs.

\item \textbf{Coordination}: Blockchain can take up coordinator functions. For example, instead of aggregating $\theta_i$ directly on-chain—which can lead to computational and storage overhead—blockchain can facilitate random leader selection through oracles
% \todo{explain it}
. Other coordination functions enabled by blockchain include trustless voting systems \cite{MH8_Fadaeddini2019, MH9_He2021, KT01_Toyoda2019BigData, KT02_Toyoda2020Access, KT11_Xuan2021SCN}, and the maintenance of essential federated learning (FL) data such as update verifications and member registries \cite{LW5_FL_HomeAppliances_IoT, MH11_Desai2021, MH1_Chai2021, KT11_Xuan2021SCN, LW10_refiner, MH13_Li2020, KT04_Zou2021WCNC, LW1_witt2021rewardbased, MH11_Desai2021, LW5_FL_HomeAppliances_IoT, LW1_witt2021rewardbased, LW10_refiner, KT01_Toyoda2019BigData, KT02_Toyoda2020Access, LW2_Weng2021, LW9_FedCoin}. These processes are managed in a transparent and immutable manner, enhancing accountability.

\item \textbf{Payment}: Incorporates cryptocurrency transactions within smart contracts for instant, automatic payments, thereby enhancing the efficiency of transaction processes within FL.

\item \textbf{Storage}: Although decentralized storage on BC is resource-intensive due to redundancy, it provides critical benefits in terms of data auditability and trust. The immutable and transparent nature of BC supports shared access and verifiability of crucial FLF data, such as machine learning models, reputation scores, user information, and votes, thereby facilitating accountability and precise reward calculations \cite{LW7_Zhang_Reputation, LW13_TowardsReputationINFOCOMM, LW2_Weng2021, LW8_Privacy_IoV,LW11_RewardResponseGame, MH1_Chai2021, Bao2019FLChain, MH4_Ma2021, MH9_He2021, MH10_Qu2021, MH11_Desai2021, KT01_Toyoda2019BigData, KT02_Toyoda2020Access, LW1_witt2021rewardbased, LW10_refiner, LW9_FedCoin}. Protocols like the interplanetary file-system (IPFS \cite{ipfs}) can be used to store data off-chain

\end{enumerate}

% KFCA is (i) lightweight, unlike Shapley value calculation; (ii) capable of computing rewards in one shot; and (iii) does not require access to extensive report distributions, unlike CA \cite{LIU_CA} or Peer Truth Serum \cite{LW1_witt2021rewardbased}. These features make KFCA a promising candidate for blockchain implementation, promoting decentralized and incentivized FL \cite{LeonSP}. The appendix (\ref{app:blockchain}) details an architecture draft of a decentralized and incentivized FL framework on the blockchain. 

\section{Decentralized and Incentivized Federated Learning: A framework}
\label{sec:FLF}

\begin{figure*}[t]
    \centering
    \includegraphics[width=\textwidth]{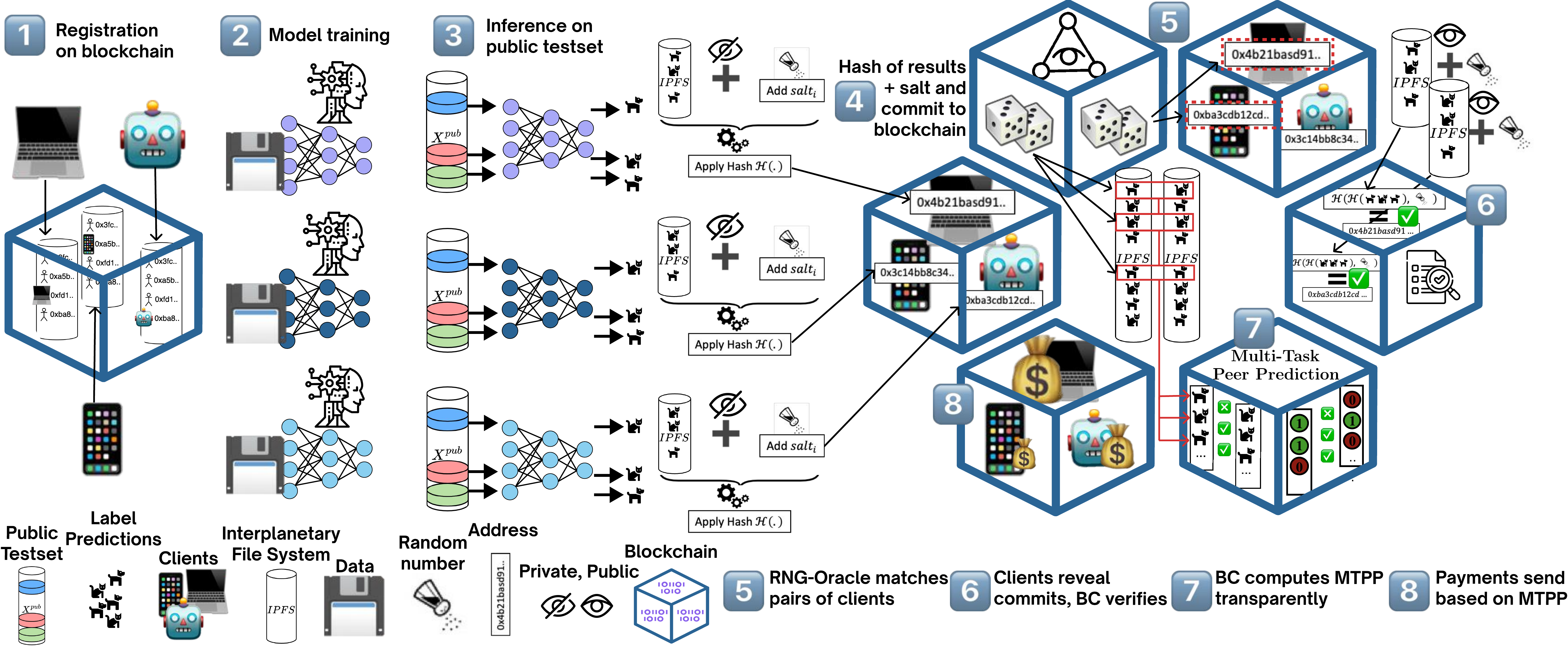}
    \caption{Incentivized and Decentralized Federated Learning: Application of a Multi-task peer prediction mechanism to incentivize FL on Blockchain.}
    \label{fig:KFCA_X_Blockchain}
\end{figure*}

Computationally intensive contribution measurement methods conflict with the strict computation and storage limits of blockchain systems because multiple blockchain nodes must store and compute smart contract transactions in parallel. 
Ideally, the incentive system should be (i) lightweight, (ii) capable of computing rewards in a single operation, and (iii) compatible with integer-operations only. MTPP - in contrast to explicit measurements like the Shapley value - could be the remedy here \cite{CA_HONGTAO, LIU_CA, LW1_witt2021rewardbased} as they base their rewards on the correlation of other reports to incentivized honest model training by the clients. 
The extent of the clients' interaction with the blockchain, whether as a communication intermediary or as participants in functions like validation, depends on the specific blockchain architecture used.

\subsection{MTPP to incentivize FL on blockchain}
MTPP have promising applications in the domain of decentralized and incentivized FL due to its simplicity in incentivizing and rewarding clients. Fig.~\ref{fig:KFCA_X_Blockchain} illustrates a possible integration of an MTPP, leaning on Correlated Agreement \cite{OriginalCA } - on Blockchain (e.g., Ethereum Virtual Machine (EVM)) in the following steps. 

\begin{enumerate}
    \item \textbf{Registration on Blockchain:} First, clients who want to participate have to register themselves on the Blockchain. On the EVM, this can happen with a smart contract, where clients register their public address and sign the registration with their private key. Depending on the use cases, clients might be required to initially commit a certain financial stake (for example, if clients are fully anonymous or have a self-interest in a trained model). Another option is that the registration can be set up such that only pre-known clients are allowed to participate.  
    \item  \textbf{Model Training} Initially, clients train their local models on their respective data by applying an appropriate optimization algorithm. The goal of this training process is to optimize a specified objective, which often involves minimizing a loss function.
    \item \textbf{Inference on Public Testset:} Given the public testset $X^{pub}$, clients predict the labels e.g.\cite{LIU_CA} or the  quantized model parameters, e.g., \cite{CA_HONGTAO}. Note that Fig. \ref{fig:KFCA_X_Blockchain} illustrates the case where the ML task is to classify an image. 

\item \textbf{Hash-Commit:} Blockchain is inherently transparent, meaning every node has access to all information on the blockchain. To perform MTPP, clients need to reveal their reports of signals to the chain so that a smart contract can calculate the rewards. However, this introduces a problem: malicious clients could wait to see what other clients have reported and then report something similar to avoid model training while still getting rewarded. To prevent this, we introduce a hash-commit scheme, similar to \cite{LW1_witt2021rewardbased}. In this scheme, instead of publishing results directly, the information is concatenated with a random number, referred to as a \textit{salt}, and then hashed. Adding the salt prevents brute-force attacks that could potentially restore the input given a hashed output, especially if the input space is limited. Once a sufficient number of clients have \textit{committed}—signing with their private key to secure the data and ensure its authenticity—the smart contract requires the clients to \textbf{reveal} both the initial information and the salt (see step 6: reveal).

Because every node in the blockchain network must duplicate all information and computation, storing large amounts of data negatively impacts the network’s scalability. To manage this issue while ensuring data remains unchangeable and secure, we use the Interplanetary File System (IPFS) \cite{benet2014ipfs}. IPFS provides a collision-proof hash for each piece of data. Instead of storing the clients’ signals directly on the blockchain, the hash acts as a unique pointer to the data, ensuring it can be securely and reliably identified without the overhead.
\item \textbf{Random Number Oracle to Randomly Match Two Clients:} A crucial step of MTPP is the random pairing of clients to compute the rewards, necessitating a verifiable randomization mechanism. We employ a decentralized Verifiable Random Function (VRF), such as Chainlink VRF \cite{ChainlinkVRF2024}. This VRF functions as an \textit{oracle} that can generate a cryptographically secure random seed. The random seed is then used to facilitate the random matching of two committed clients. Utilizing a decentralized VRF in this process ensures the generation of unbiased and cryptographically verifiable randomness, thereby upholding the integrity and fairness of the client pairing process in the MTPP algorithm.

   \item \textbf{Reveal:} Following the random selection of clients by the oracle, the clients must now reveal their previously committed information. The selected clients are required to disclose two key pieces of information to the blockchain: the IPFS pointer to their labels and their respective individual \textit{salt} values. The blockchain then hashes both the IPFS pointer and the salt for each client. This hashed combination is compared against the initial commit made by the clients to verify the authenticity and integrity of the information, ensuring that the data provided by the clients has not been altered since the initial commitment.

\item \textbf{MTPP:} Similar to step 5, the Verifiable Random Function (VRF) is used to generate a random seed for selecting random samples. Once this selection is completed, a smart contract on the blockchain computes the reward scores for the clients using the MTPP algorithm.

\item \textbf{Payment:} Utilizing the cryptocurrency features of blockchain platforms, the rewards are adjusted according to the predefined rules of the protocol. Payments are then automatically and immediately transferred to the clients’ accounts, ideally occurring in real-time as specified by the MTPP details.

\end{enumerate}

\section{Limitations and Future research}

The proposed concept is a high-level architecture and requires further specification particularly of a blockchain-compatible MTPP, in order to be implemented in practice. Here, the following limitations have to be overcome and investigated in future research:

\begin{enumerate}

\item \textbf{MTPP Specifics:} The problem with Correlated Agreement (CA) and Peer Truth Serum (PTS) is that full information on the data distribution is needed, which requires an overview. Additionally, calculating the delta-matrix for CA still requires significant computation and cannot be done on the fly.
\item \textbf{Implementation and Performance Tests:} The performance of such a system heavily depends on the underlying blockchain architecture, particularly whether the consensus mechanism causes overhead, how the data  structure influences data queries (Merkle-root based data structures used in IPFS might be slow), the number of peers in the network, and the encoding and data-type specifics of the underlying virtual machine. All of these factors contribute to computational, storage, and complexity overhead, a trade-off for a decentralized and incentivized FL paradigm.
\item \textbf{Random Validator Selection:} The diagram does not specify who should aggregate the respective model parameters. A suggestion would be that the oracle randomly chooses a participant as a parameter - perhaps in an optimistic fashion \cite{conway2024opml}
or as a majority vote \cite{LW1_witt2021rewardbased}. 
\end{enumerate}

\section{Conclusion}
This work introduces a concept to decentralize and incentivize Federated Learning by leveraging Multi-Task Peer Prediction to reward informed and truthful AI training. By integrating general-purpose blockchain technology, this framework achieves genuine decentralization, setting the stage for truly democratic AI training processes.
\bibliographystyle{IEEEtran}
\bibliography{bibliography}

\end{document}